\documentclass[12pt]{iopart}

\usepackage{graphicx}
\begin{document}

\title{Least momentum space frustration as a condition 
for ``high $T_c$ sweet spot''  in the iron-based superconductors}

\author{Hidetomo Usui$^1$, Katsuhiro Suzuki$^{1,2}$, 
and Kazuhiko Kuroki$^{1,2}$}

\address{$^1$Department of Engineering Science, The University of Electro-Communications, Chofu, Tokyo 182-8585, Japan \\
$^2$JST, TRIP, Chiyoda, Tokyo 102-0075, Japan}
\ead{kuroki@vivace.e-one.uec.ac.jp}

\begin{abstract}
In the present paper, we describe how the band structure and the 
Fermi surface of the iron-based superconductors vary as the 
Fe-As-Fe bond angle changes. We discuss how these Fermi surface 
configurations affect the superconductivity mediated by spin 
fluctuations, and show that in several situations, frustration in the 
sign of the gap function arises due to the repulsive pairing 
interactions that requires sign change of the order parameter.
Such a frustration can result in nodes or very small gaps, and 
generally works destructively against superconductivity. 
Conversely, we propose that the optimal condition for superconductivity 
is realized for the Fermi surface configuration that gives the 
least frustration while maximizing the Fermi surface 
multiplicity. This is realized when there are three hole 
Fermi surfaces, where two of them have $d_{XZ/YZ}$ orbital 
character and one has $d_{X^2-Y^2}$ {\it for all $k_z$} in the 
three dimensional Brillouin zone. Looking at the band structures of 
various iron-based superconductors, the 
occurrence of such a ``sweet spot'' situation is limited to  a narrow 
window.
\end{abstract}

\maketitle

\section{Introduction}
The discovery of high temperature superconductivity in the 
iron-based superconductors\cite{Hosono} has attracted much attention 
in many aspects.
Not only the high $T_c$ itself, but also a number of experiments 
indicating non-universality of the superconducting gap function, 
such as sign reversing, anisotropy, or the presence of nodes, 
suggest an unconventional pairing mechanism. 
Most probable candidate for such an unconventional mechanism 
is the pairing mediated by 
spin fluctuations, where the superconducting gap changes  
sign between the disconnected Fermi surfaces, namely,  
the so-called $s\pm$ pairing\cite{Mazin,Kuroki1st}. 

Back in 2001, one of the present authors proposed that 
spin fluctuation mediated pairing in systems with nested 
disconnected Fermi surfaces may give rise to a very high $T_c$ 
superconductivity\cite{KA,KA2}. 
The idea is that the repulsive pairing interaction 
mediated by spin fluctuations can be fully exploited without 
introducing nodes of the superconducting gap on the Fermi surfaces.
Although the Fermi surface of the iron-based superconductors 
does resemble the proposed Fermi surface configuration, there are 
some important differences. One is that the Fermi surface in the 
iron-based superconductors has multiple orbital characters, and 
second is that there can be frustrations in the sign of the gap 
function, which can give rise to gap nodes on the Fermi surface. 
In the present paper, we focus on this gap-sign frustration problem, 
and propose that the Fermi surface configuration that gives the 
least frustration provides the optimal condition for high $T_c$ 
in the iron-based superconductors. In ref.\cite{Usui}, two of the 
present authors pointed out that maximizing the Fermi surface 
multiplicity leads to the optimization for $T_c$, but considering 
the frustration problem studied in the present paper, the 
``sweet spot'' for high $T_c$ is further limited to a narrow window 
of the Fermi surface configuration.

\section{Typical band structure and Fermi surfaces}
\begin{figure}
\begin{center}
\includegraphics[width=8cm,clip]{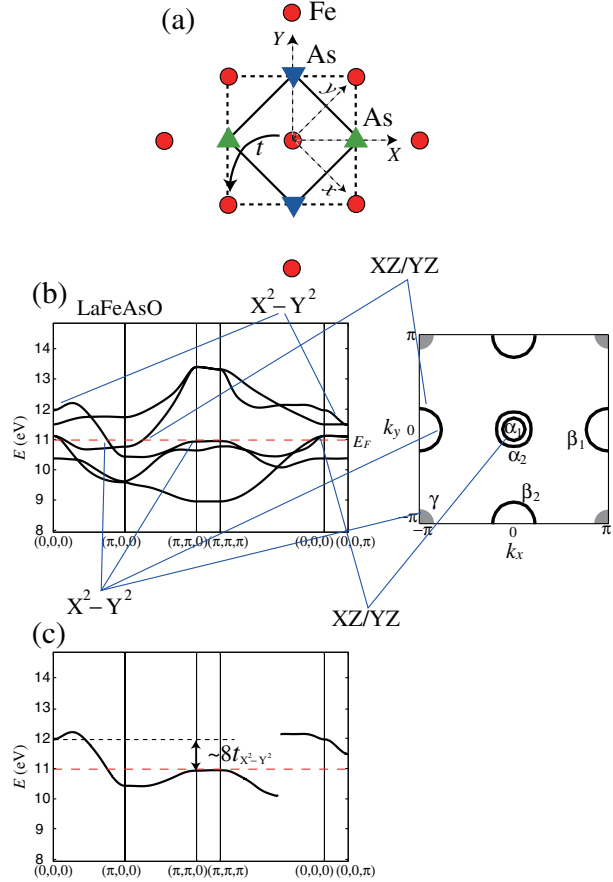}
\end{center}
\caption{(a) The original (dashed lines) and 
reduced (solid) unit cells with $\bullet$ (Fe), $\nabla$ (As 
below the Fe plane) and $\triangle$ (above Fe). 
(b) The band structure (left) of the five-band model for LaFeAsO, and the 
Fermi surface (right) at $k_z=0$ for 10 percent electron doping. 
The main orbital 
characters of some portions of the bands and the Fermi surface 
are indicated.  
The dashed horizontal line in the band structure indicates the Fermi level 
for 10 percent electron doping. 
The short arrow in the band structure indicates the 
position of the Dirac cone closest to the Fermi level.  
The gray areas in the Fermi surface 
around the zone corners represent the $\gamma$  Fermi surface, 
which is barely absent for 10 percent electron doping.
(c) The portion of the band that has mainly the $d_{X^2-Y^2}$ 
orbital character.
\label{fig1}}
\end{figure}

Let us first describe the band structure of LaFeAsO.
LaFeAsO takes a layered structure, where Fe atoms form a 
square lattice in each layer, 
sandwiched by As atoms with tetrahedral coordination.
We use the  band structure obtained from first principles\cite{pwscf}
to construct the maximally localized 
Wannier functions\cite{MaxLoc}. 
These Wannier functions 
have five orbital symmetries ($d_{3Z^2-R^2}$, 
$d_{XZ}$, $d_{YZ}$, $d_{X^2-Y^2}$, 
$d_{XY}$), where $X, Y, Z$ refer to those for this unit cell 
with two Fe sites as shown in Fig.\ref{fig1}(a). 
The two Wannier orbitals in 
each unit cell are equivalent in that each Fe atom has the same 
local arrangement of other atoms.
We can then take a unit cell that 
contains only one orbital per symmetry by 
unfolding the Brillouin zone,
and an effective five-band model on a 
square lattice is obtained, where 
$x$ and $y$ axes are rotated by 
45 degrees from $X$-$Y$.

In Fig.\ref{fig1}(b) (right), 
the Fermi surface for 10 percent electron doping 
is shown in the two-dimensional unfolded Brillouin zone.
The Fermi surface consists of four pieces:   
two concentric hole pockets (denoted here as $\alpha_1$, $\alpha_2$) 
centered around $(k_x, k_y)=(0,0)$, two electron pockets 
around $(\pi,0)$ $(\beta_1)$ or $(0,\pi)$ $(\beta_2)$, respectively. 
Besides these pieces of the Fermi surface, there is a portion of the band 
near $(\pi,\pi)$ that 
touches the $E_F$, so that 
the portion acts as a ``quasi Fermi surface $(\gamma)$'' around $(\pi,\pi)$. 
As for the orbital character, $\alpha$ and portions of $\beta$ near 
Brillouin zone edge have mainly $d_{XZ}$ and $d_{YZ}$ character, 
while the portions of 
$\beta$ away from the Brillouin zone edge and $\gamma$ have 
mainly $d_{X^2-Y^2}$ orbital character (see also Fig.\ref{fig1}(c)).

\section{Fermi surface appearance/disappearance against the bond angle}
\label{FSvsangle}

\begin{figure}
\begin{center}
\includegraphics[width=8cm,clip]{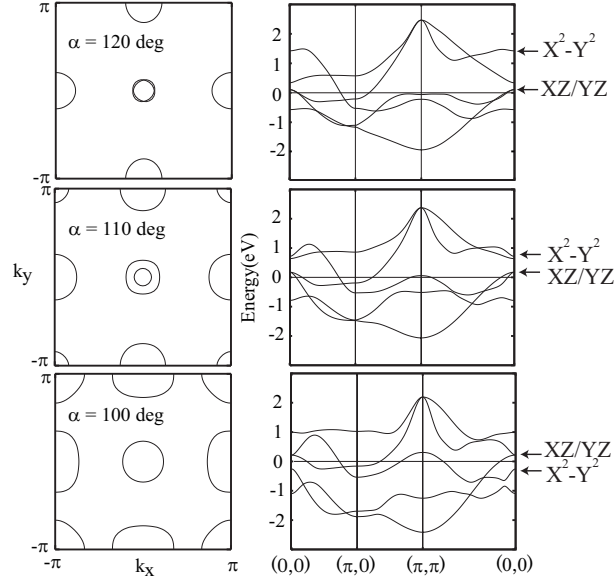}
\end{center}
\caption{ The band structure and the Fermi surface of hypothetical lattice 
structures of LaFeAsO with varying the Fe-As-Fe bond angle 
while fixing the Fe-As bond length to its original value. 
Position of the bands having $X^2-Y^2$ and $XZ/YZ$ orbital characters 
at (0,0) is indicated.}
\label{fig2}
\end{figure}

The band structure and the Fermi surfaces of the iron-based 
superconductors are sensitive to the 
lattice structure. In this section, we consider 
hypothetical lattice structures of LaFeAsO, where 
we fix the bond length at its original length
and vary the bond angle $\alpha$(Fig.\ref{fig2}(a))\cite{Usui}.
We first neglect the three dimensionality (out-of-plane hoppings), and 
consider a two dimensional model.
The Fermi surface is obtained for 10 percent electron doping.
When the bond angle is large, two hole Fermi surfaces, 
$\alpha_1$  and $\alpha_2$ 
are present around the wave vector (0,0), while the 
$\gamma$ around $(\pi,\pi)$ is missing. As $\alpha$ decreases, 
the $\gamma$ Fermi surface appears around $(\pi,\pi)$, and there are now 
three hole Fermi surfaces. This appearance of the additional Fermi surface 
has been noticed as an effect of increasing the pnictogen height
\cite{Singh,Vildosola,KKprb,Lebegue,OKAndersen}.
When $\alpha$ is further reduced, the $\alpha_1$ Fermi surface 
disappears, and again the 
Fermi surface multiplicity reduces to two, but in this 
case one around $(0,0)$ and another around $(\pi,\pi)$.
Such a disappearance of the $\alpha_1$ hole Fermi surface was 
first noticed in the band calculation of Ca$_2$Al$_4$O$_6$Fe$_2$As$_2$
\cite{Shirage} by Miyake {\it et al.} in refs.\cite{Miyake,Kosugi}.
This material indeed has very small bond angle of about 102 degrees.

\begin{figure}
\begin{center}
\includegraphics[width=8cm,clip]{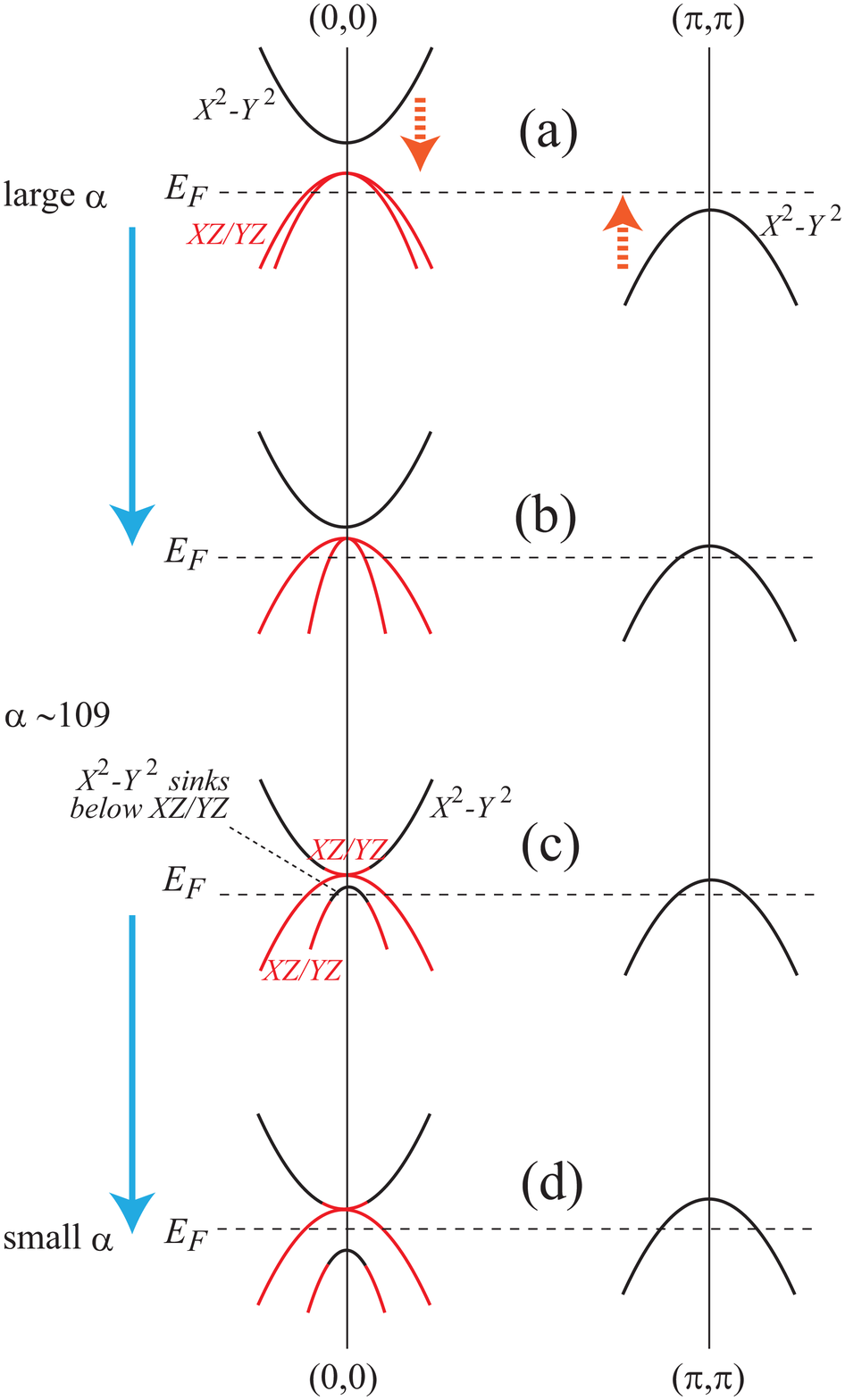}
\end{center}
\caption{A schematic figure for the band structure variation against 
the bond angle $\alpha$. The black (red) portions indicate the bands 
with strong $X^2-Y^2$ $(XZ/YZ)$ orbital character.}
\label{fig3}
\end{figure}

Fig.\ref{fig3} explains schematically the band structure/Fermi surface 
variation against the 
bond angle reduction\cite{Miyake}. 
As the bond angle is reduced, the $X^2-Y^2$ band below 
the Fermi level around $(\pi,\pi)$ rises up, while another $X^2-Y^2$ portion 
of the band above the Fermi level around $(0,0)$ comes down. 
This band deformation can be understood from Fig.\ref{fig1}(c), where the 
$X^2-Y^2$ portion of the band for the original LaFeAsO is extracted. 
In the tightbinding picture, the energy difference between 
the wave vectors $(0,0)$ and $(\pi,\pi)$ is roughly equal to $8t_{X^2-Y^2}$, 
where $t_{X^2-Y^2}$ 
is the nearest neighbor hopping of the $X^2-Y^2$ orbital. As the bond angle 
is reduced, the contribution to $t_{X^2-Y^2}$ from the 
Fe$\rightarrow$As$\rightarrow$Fe path decreases, while that from the direct 
Fe$\rightarrow$ Fe path increases. The two contributions have opposite signs,
so that the reduction of the bond angle results 
in a decrease of $t_{X^2-Y^2}$\cite{Miyake}.

When the bottom of the upper $X^2-Y^2$ portion 
sinks below the $XZ/YZ$ bands around $(0,0)$, 
a band structure reconstruction takes place, and now the two bands that are 
degenerate at $(0,0)$ repel with 
each other (one going up, the other going down)
as the wave vector moves away from $(0,0)$ (Fig.\ref{fig3}(c)). 
In this situation, the band below these two bands
has $X^2-Y^2$ character near $(0,0)$, and changes its character to 
$XZ/YZ$ as the wave number increases. Therefore, just before the 
inner hole ($\alpha_1$) Fermi surface disappears, 
it has strong $X^2-Y^2$ character.

\section{Frustration in the superconducting gap 
in the absence of $\gamma$ Fermi surface : a brief review}

Having understood the Fermi surface variation against the bond angle, 
we now consider how this should affect the gap function of the 
superconductivity mediated by spin fluctuations. 
The pairing interaction mediated by spin fluctuations is repulsive, 
so the superconducting gap has an tendency of changing its sign 
between the initial and final wave vectors of the pair scattering.
Another important point is that the spin fluctuations develop at 
wave vectors that bridge the portions of the Fermi surface having the 
same orbital character\cite{KKprb}. 

In the presence of the $X^2-Y^2$ originated $\gamma$ and $XZ/YZ$ originated 
$\alpha_1$ and $\alpha_2$ Fermi surfaces (case (b) in Fig.\ref{fig3}) ,
it is known that the $\gamma$-$\beta$ interaction among portions having 
$X^2-Y^2$ character and the $\alpha$-$\beta$ interactions among portions with 
$XZ/YZ$ character dominate, and the superconducting gap is fully open on 
all the Fermi surfaces while changing its sign as $+$, $-$, $+$ along 
$\alpha$, $\beta$, $\gamma$ Fermi surfaces. This is the fully gapped 
$s\pm$ gap\cite{KKprb,DHLee,Thomale}.

\begin{figure}
\begin{center}
\includegraphics[width=10cm,clip]{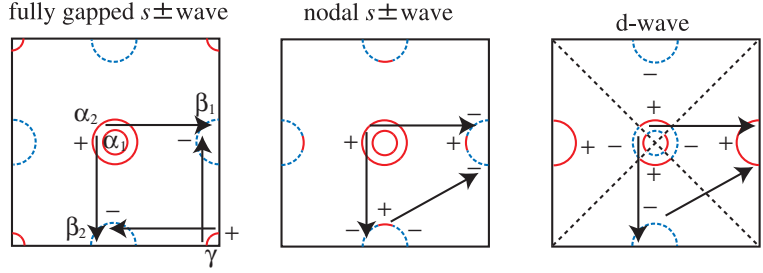}
\end{center}
\caption{The fully-gapped $s\pm$ wave, the nodal 
$s\pm$ wave and $d$ wave gap 
are schematically shown. The solid red (dashed blue) curves 
represent positive (negative) sign of the gap. The arrows 
indicate the dominating nesting vectors.}
\label{fig4}
\end{figure}

On the other hand, in the absence of the $\gamma$ Fermi surface 
(case (a) in Fig.\ref{fig3}), the pairing 
interaction between the 
$X^2-Y^2$ portions of the $\beta$ Fermi surfaces and that 
between the $XZ/YZ$ portions of the $\alpha$ and $\beta$ Fermi surfaces 
results in a frustration in the sign of the gap function. 
This situation was studied 
in several previous papers\cite{Graser,KKprb,DHLee,Thomale}. 
This can result in 
either nodal $s\pm$-wave or $d$-wave pairings, where the nodes of the gap 
go into the $\beta$ Fermi surface in the former, and $\alpha$ in the latter.
A schematic figure summarizing the above  
is shown in Fig.\ref{fig4}\cite{KKprb}.
As was shown in ref.\cite{KKprb}, this frustration effect degrades 
$T_c$ of the superconductivity, so that the lattice structure acts as 
a switch between high $T_c$ nodeless and low $T_c$ nodal superconductivity.

\section{Frustration in the case of nearly vanishing $\alpha_1$ Fermi surface}

\begin{figure}
\begin{center}
\includegraphics[width=8cm,clip]{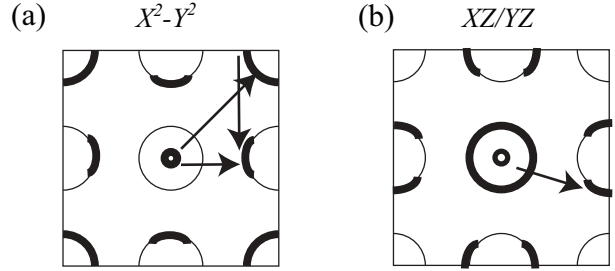}
\end{center}
\caption{Dominating pairing interactions for the (a)$X^2-Y^2$ and (b) $XZ/YZ$
portions of the Fermi surface in the case where the inner hole Fermi surface 
($\alpha_1$) is barely present. In this case, $\alpha_1$ is a mixture of 
$X^2-Y^2$ and $XZ/YZ$.}
\label{fig5}
\end{figure}

Here we discuss another situation where the frustration arises in the 
sign of the superconducting gap function. 
As mentioned in section\ref{FSvsangle}, 
 the $X^2-Y^2$ orbital character strongly 
mixes into the $\alpha_1$ Fermi surface 
just before the Fermi surface vanishes as the 
bond angle is reduced. In this situation, there are now $X^2-Y^2$ 
components on $\alpha_1$, $\beta$, and $\gamma$ (if present) Fermi surfaces.
Since these Fermi surfaces interact with repulsive pairing interactions,
once again, a frustration arises in the sign of the superconducting gap 
(Fig.\ref{fig5}(a)). 
In addition to this, there can also be 
some $XZ/YZ$ component remaining in the $\alpha_1$ Fermi surface, 
and this portion tends to change the sign from the $\beta$ Fermi 
surfaces (Fig.\ref{fig5}(b)) so this can be another factor for the 
frustration. 

\begin{figure}
\begin{center}
\includegraphics[width=12cm,clip]{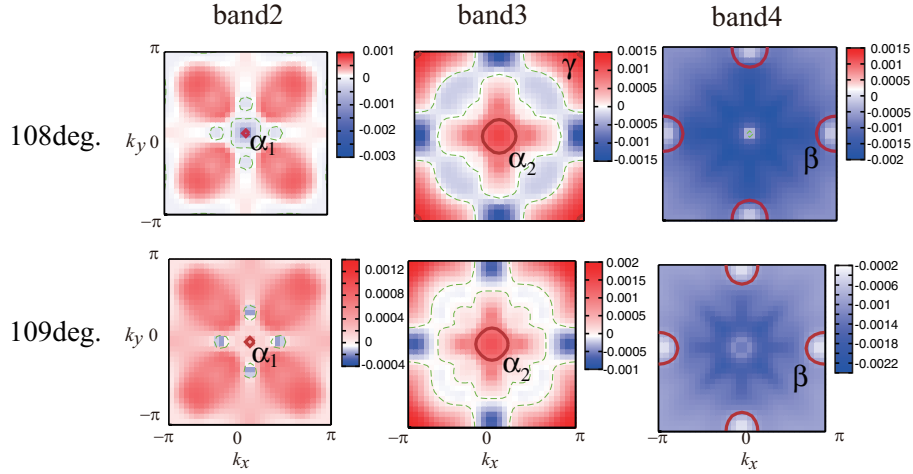}
\end{center}
\caption{The gap function obtained by FLEX for the 
hypothetical lattice structures of Ca$_4$Al$_2$O$_6$Fe$_2$As$_2$.
The bond angle $\alpha$ is varied to 108 or 109 degrees, while the 
bond length is fixed to the original value.}
\label{fig6}
\end{figure}

To actually see this frustration effect, 
here we consider 
hypothetical lattice structures of 
Ca$_4$Al$_2$O$_6$Fe$_2$As$_2$\cite{Shirage}, 
where we vary the bond angle from 108 to 109 degrees, 
and contrust a two dimensional five orbital model in the unfolded 
Brillouin zone.
In this angle regime, the $\alpha_1$ Fermi surface is barely 
present, and it is 
indeed composed of mixed $X^2-Y^2$ and $XZ/YZ$ orbital components.
We apply fluctuation exchange approximation to this model\cite{Bickers}, 
and obtain the eigenfunction (gap function) of the linearized  
Eliashberg equation as was done in ref.\cite{Usui}.
In Fig.\ref{fig6}, we show the gap function for the 
two angles for 10 percent electron doping and temperature $T=0.01$eV.  
It can be seen that the magnitude of the gap on the 
$\alpha_1$ Fermi surface is very small, and its sign actually changes as the 
bond angle is varied, reflecting the frustration.

\section{Effect of three dimensionality}

\begin{figure}
\begin{center}
\includegraphics[width=10cm,clip]{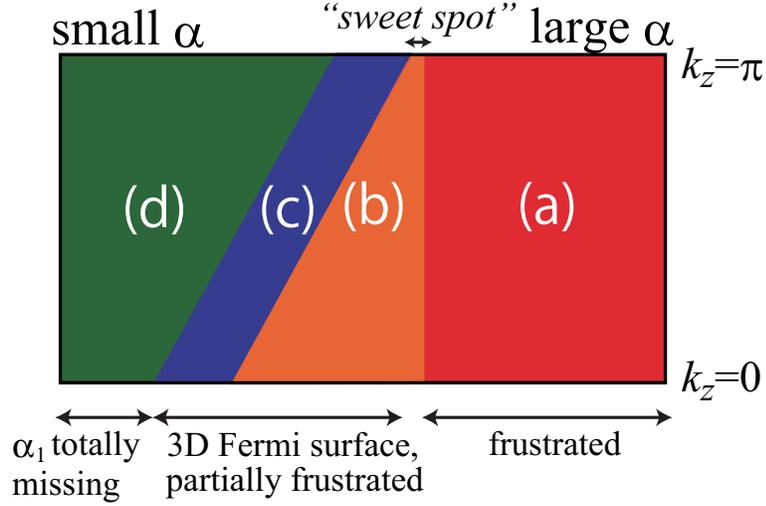}
\end{center}
\caption{A schematic figure representing the band structure and 
the hole Fermi surface variation against the bond angle 
for moderately three dimensional systems. (a) to (d) correspond to 
the configurations shown in Fig.\ref{fig3}. The ``sweet spot'' 
is restricted to the regime where the $X^2-Y^2$ originated 
$\gamma$ Fermi surface 
is effective and the $\alpha_1$ Fermi surface has $XZ/YZ$ character 
for all $k_z$. }
\label{fig7}
\end{figure}

When the systems exhibit some three dimensionality , the 
above mentioned variation of the Fermi surface configuration 
against the bond angle depends on $k_z$. This is shown 
schematically in Fig.\ref{fig7}. In systems with moderate three 
dimensionality, the orbital character change of the $\alpha_1$ 
Fermi surface and its disappearance first starts at $k_z=\pi$ 
(around Z point)  
as the bond angle is reduced, and ends at $k_z=0$ ($\Gamma$ point). 
Namely, the 
$\alpha_1$ Fermi surface becomes three dimensional before it 
disappears completely. In this case, 
the above mentioned frustration effect should be present 
around the top and the bottom portions of the three dimensional 
Fermi surface.

\begin{figure}
\begin{center}
\includegraphics[width=10cm,clip]{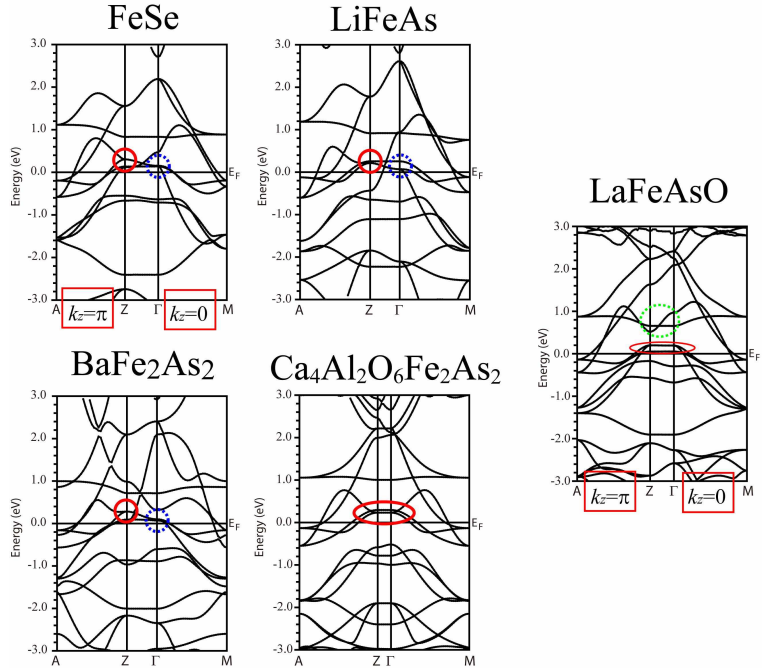}
\end{center}
\caption{First principles band structure for various iron-based 
superconductors. $k_z=0$ and $k_z=\pi$ planes are shown. 
The red solid circles (or elipse) indicates the portion 
where the two bands degenerate at wave vector $(k_x,k_y)=(0,0)$ 
repel with each other as the wave number increases, while the 
dashed blue cirles are the portions where the two bands degenerate 
at (0,0) both form a hole Fermi surface. For LaFeAsO, a typical 
example for 1111 systems, the three dimensional band  along $\Gamma$-Z
having $X^2-Y^2$ character (green dashed elipse) 
stays above the degenerate $XZ/YZ$ bands (red solid).}
\label{fig8}
\end{figure}

In fact, the above mentioned three dimensionality of the band structure  is 
rather common for the iron-based superconductors. 
This has been discussed in detail in ref.\cite{OKAndersen}.
In Fig.\ref{fig8}, we show the band structure of various 
iron-based superconductors for $k_z=0$ and $k_z=\pi$ planes
(in the original folded Brillouin zone) calculated by 
using the Wien2k package\cite{Wien2k}. One can see that, except for 
LaFeAsO, the two bands degenerate at Z point $(0,0,\pi)$ 
repel with each other 
as the wave vector moves away from Z in the $k_z=\pi$ plane. 
This means that configurations (c) or (d) in Fig.\ref{fig3} 
are realized for a certain range of $k_z$. On the other hand, 
in FeSe, LiFeAs and  BaFe$_2$As$_2$ the band structure near the 
$k_z=0$ plane takes configuration (b), so that in these materials, 
the band structure near the Fermi level is indeed three dimensional.

\begin{figure}
\begin{center}
\includegraphics[width=10cm,clip]{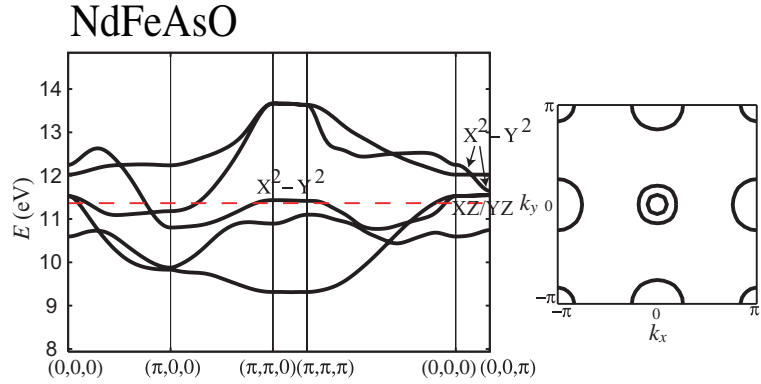}
\end{center}
\caption{The band structure of NdFeAsO in the unfolded Brillouin zone.
The Fermi level is for 10 percent electron doping. 
The two conditions for least 
frustration is satisfied: (i) the $X^2-Y^2$ Fermi surface is present 
around the wave vector $(\pi,\pi)$, and (ii) the three dimensional 
$X^2-Y^2$ band 
 from $(0,0,0)$ to $(0,0,\pi)$ does not 
intersect the $XZ/YZ$ bands for all $k_z$.}
\label{fig9}
\end{figure}

Conversely, the 1111 systems can be considered as somewhat exceptional 
in that, although the three dimensional $X^2-Y^2$ band does exist, 
it does not come down too rapidly before the $X^2-Y^2$ originated $\gamma$ 
Fermi surface appears. In fact, NdFeAsO, one of the 
materials having the highest $T_c$, seems to have the least frustration 
from the above mentioned viewpoint. Namely, as seen in the band structure 
 shown in Fig.\ref{fig9} in the unfolded Brillouin zone\cite{KKprb}, 
the $\gamma$ 
Fermi surface is present but the three dimensional $X^2-Y^2$ band 
still lies above the $XZ/YZ$ bands for all $k_z$. 
On the other hand, from the comparison to the band structure of LaFeAsO
(Fig.\ref{fig1}(b)), 
it can be seen that 
as a trade-off for the appearance of the $\gamma$ 
Fermi surface around $(\pi,\pi)$, the three dimensional $X^2-Y^2$ band 
along $(0,0,0)$-$(0,0,\pi)$ 
has certainly come down very close to the $XZ/YZ$ band, 
and for smaller bond angle or higher pnictogen position, 
the band reconstruction and thus the frustration starts to take place.
Therefore, from the present viewpoint, the ``sweet spot'' for high $T_c$ 
is restricted to a narrow window, 
which may explain the experimental observations\cite{Lee,Takano}.

\section{Conclusion}
As described in the present paper, the multiplicity of the 
Fermi surfaces and even their orbital characters change   
as the lattice structure is varied. Because of the presence of 
multiple Fermi surfaces and the repulsive pairing interaction 
mediated by the spin fluctuations, frustration can arise in the 
sign of the gap on the disconnected Fermi surfaces. This 
can result in nodal structures in the gap function, and should generally 
work destructively against superconductivity. Conversely, from the present 
viewpoint, the optimal situation 
for the spin fluctuation mediated superconductivity 
is when the $X^2-Y^2$ $\gamma$ Fermi surface is effective and the $\alpha_1$ 
Fermi surface has $XZ/YZ$ orbital character for {\it all} $k_z$. 
As far as the first principles band calculations are concerned, 
this situation is realized in limited materials like 
NdFeAsO and SmFeAsO, where the 
highest $T_c$ among the iron based superconductors is observed experimentally. 
From this viewpoint, further band structure calculation studies 
may give useful information 
for obtaining new related superconductors with higher $T_c$.

\section*{Acknowledgment}
We acknowledge C.H. Lee, 
T. Miyake, O.K. Andersen, H. Mukuda, H. Kinouchi, and Y. Kitaoka  
for illuminating discussions.

\end{document}